\documentclass{PoS}

\newcommand{\Fig}[1]{Fig.~\protect\ref{#1}}

\newlength{\Tatescale}
\newlength{\figwidth}
\newcommand{\Cut}[1]{}
 
\newcommand{\beq}{\begin{eqnarray}}
\newcommand{\eeq}{\end{eqnarray}}

\newcommand{\bL}{\mbox{\boldmath $L$}}

\newcommand{\bS}{\mbox{\boldmath $S$}}

\newcommand{\bsigma}{\mbox{\boldmath $\sigma$}}
\newcommand{\btau}{\mbox{\boldmath $\tau$}}

\newcommand{\pslash}{p\kern-1ex /}
\newcommand{\kslash}{k\kern-1ex /}
\newcommand{\qslash}{q\kern-1ex /}
\newcommand{\lslash}{l\kern-1ex /}
\newcommand{\sslash}{s\kern-1ex /}
\newcommand{\paslash}{p_a\kern-2ex /}
\newcommand{\pbslash}{p_b\kern-2ex /}
\newcommand{\Dslash}{{\cal D}\kern-1.5ex /}
\newcommand{\dslash}{\partial\kern-1.2ex /}

\title{Nuclear Forces from Lattice QCD}

\ShortTitle{Nuclear Forces from Lattice QCD}
\author{\speaker{Tetsuo Hatsuda} \ \ (for HAL QCD Collaboration) \thanks{
This research  was supported  in part by the
Grant-in-Aid for Scientific Research on Innovative Areas (No. 2004: 20105003).}
\hspace{2cm} \includegraphics[width=4.0cm]{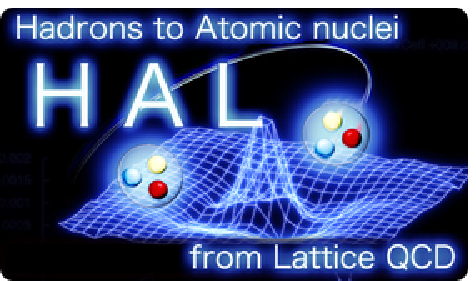} \\
        Phys. Dep., Univ. of Tokyo, Tokyo 113-0033, Japan\\   
        E-mail: \email{hatsuda@phys.s.u-tokyo.ac.jp}\\
  \\
  }

\abstract{A method to extract 
  nucleon-nucleon (NN) potentials from  the Bethe-Salpeter amplitude  in
  lattice QCD is presented.  It is applied to the two nucleons on the lattice with  
   quenched  QCD simulations. By disentangling the mixing between the S-state and
  the  D-state,  we obtain central and tensor potentials 
 in the leading order  of the velocity expansion of the non-local NN potential.
  The  spatial structure, 
   the quark mass dependence and the velocity dependence of 
   the NN potential are  analyzed.  Preliminary result in (2+1)-flavor 
   QCD simulations is also shown.
}

\FullConference{6th International Workshop on Chiral Dynamics, CD09\\
		July 6-10, 2009\\
		Bern, Switzerland}

\begin{document}

\section{Introduction}
\label{sec:intro}

 The origin of the nuclear force is one of the 
 major unsolved problems in particle and nuclear physics.
   To describe the  elastic nucleon-nucleon (NN) scattering at low-energies
 and  the deuteron properties,
 the notion of the NN potential turns out to be very useful \cite{NN-review}.
 The phenomenological  NN potentials 
 which can fit the NN data precisely 
 are known to have the following properties:
 (i)  The long range part (the relative distance 
 $r >  2$  fm) 
 is dominated by the one-pion exchange introduced by Yukawa \cite{yukawa}.
(ii)  The medium range part ($1\ {\rm fm} < r < 2$ fm) receives 
  significant contributions from the exchange of  
  two-pions ($\pi\pi$) and heavy mesons ($\rho$, $\omega$, and $\sigma$).
(iii)  The short  range part ($r < 1$ fm) is best described by
  a strong repulsive core as  introduced by Jastrow \cite{jastrow}.
(iv) A strong attractive spin-orbit force in the 
  isospin 1 channel exists at medium and
 short distances. (i) is related to the tensor force which is a key for
 the deuteron binding, (ii) is important for the binding of nuclei with 
 more than 2 nucleons, (iii) is important for the stability of  
 nuclei and   neutron stars, and (iv) is related to the $^3 {\rm P}_2$ neutron pairing
which leads to the neutron  superfluidity inside neutron stars \cite{VJ}.
   
 A repulsive core  surrounded by an attractive well as seen in the phenomenological 
 nuclear force
   is  a
  common feature of the ``effective" potentials between
 composite particles.  The Lenard-Jones potential between 
  neutral atoms or molecules is a well-known example in 
  atomic physics. The potential between $^4$He nuclei 
  is a typical example in nuclear physics.
  The origin of the repulsive cores in these examples
   are known to be the Pauli exclusion among electrons or among nucleons.
  The same  idea, however, is not applicable to the NN potential,
  because the quark has not only spin and  flavor
  but also color which allows 
  six quarks occupy the same state without violating  the Pauli principle. 
  To account for the repulsive core of the NN force, therefore, 
  various different ideas have been proposed so far \cite{QQ_review}:
  an exchange of the neutral
  $\omega$ meson as proposed by Nambu \cite{nambu57},  exchanges of 
   non-linear pion field, a combination of the Pauli principle
    with the one-gluon-exchange between quarks and so on.
  Despite all these efforts, convincing account of the 
  nuclear force has not yet been obtained.

\section{NN interactions from lattice QCD}

  Under the situation mentioned above, it is highly desirable to
  study the NN interactions from the first principle
  lattice QCD simulations. A theoretical framework
  suitable for such purpose was first proposed 
  by L\"{u}scher \cite{luescher}: For two hadrons in a finite $L^3$
  box,  
  an exact relation between  the energy spectrum in the box
  and the elastic scattering phase shift  was
   derived: If the range of the hadron interaction $R$  is sufficiently
  smaller than the size of the box $R<L/2$, the behavior of the 
  Bethe-Salpeter (BS)
  wave function $\psi (\vec{r})$ in the interval $R < | \vec{r} | < L/2 $
  under the periodic boundary conditions
  has sufficient information to relate the phase shift and the 
  two-particle spectrum. 
   The L\"{u}scher's method bypasses
  the difficulty to treat the real-time scattering process
  on the Euclidean 
  lattice. 
   Furthermore, it utilizes the
   finiteness of the lattice box effectively to extract the
  information of the on-shell scattering matrix and the 
  phase shift.  This approach has been applied to the NN scattering lengths
  in \cite{Fukugita:1994ve}.
  
  Recently,  we have proposed a closely related but an
   alternative
  approach to the NN interactions from lattice QCD \cite{Ishii:2006ec,Aoki:2008hh}.  
  The starting point is the
  same BS wave function  $\psi (\vec{r})$:  
  Instead of looking at the wave function 
  outside the range of the interaction,
  we consider the internal region $ |\vec{r} | < R$ and
  define an energy-independent  non-local potential $U(\vec{r}, \vec{r}')$
  from $\psi (\vec{r})$ so that  it   
  obeys the 
  Schr\"{o}dinger type equation in a finite box.
  Since $U(\vec{r}, \vec{r}')$ for strong interaction
  is localized in its spatial coordinates due to confinement
   of quarks and gluons, the potential receives
   finite volume effect only weakly in a large box. Therefore, 
   once $U$ is determined and is appropriately extrapolated to 
   $L \rightarrow \infty$, one may simply use the Schr\"{o}dinger
   equation in the infinite space to calculate the scattering phase shifts
    and bound state spectra to compare  with experimental data.    
  Further advantage of utilizing the potential is that it would be a smooth
   function of the quark masses so that it is relatively easy to handle.
    This is in sharp contrast to the 
   the scattering length  which shows a singular 
  behavior around the quark mass corresponding to the 
  formation of the NN bound state \cite{Kuramashi:1995sc}. 

  Since we consider the non-asymptotic region ($ |\vec{r} | < R$)  
  of the wave function, the resultant potential $U$ and the $T$-matrix
  are  off-shell.  Therefore, they
   depend on the nucleon interpolating operator adopted to define
   the BS wave function.  This is in a sense  an advantage, since 
    one can establish a one-to-one correspondence between the nucleon
   interpolating operator and the NN potential in QCD, which is not attainable
   in phenomenological NN potentials.  It also implies that 
  the NN potential on the lattice and the phenomenological NN potentials
  are equivalent only in the sense that they give the same phase shifts,
  so that the comparison of their spatial structures should be made only qualitatively.

\section{Non-local potential from the BS wave function}
\label{sec:BS-wave-spin_1/2}

 Let us consider the following BS wave function
 for the 6-quark state with total energy $W$ and the total three-momentum
  $\vec{P}=\vec{0}$
in a finite box; 
$\Psi_{\alpha \beta}(\vec{r}=\vec{x}-\vec{y},t)  = 
\langle {\rm vac} \vert  {n}_{\beta}(\vec{y},t) {p}_{\alpha}(\vec{x},t) 
\vert W,\vec{P}=\vec{0} 
\rangle \equiv \psi_{\alpha \beta}(\vec{r}) e^{-iWt}$.
The   local composite operators for the proton and the neutron
 are denoted by   $p_{\alpha}(\vec{x},t)$ and  $n_{\beta}(\vec{y},t)$
 with spinor indices $\alpha$ and $\beta$.
 The
 state $| W,\vec{P}=\vec{0}  \rangle $ is a
 QCD eigenstate with baryon number 2 and with the same quantum numbers as the pn system.
  One should keep in mind  that $| W,\vec{P}=\vec{0}  \rangle $ is {\it not} 
 a simple superposition of a product state $| {\rm p} \rangle \otimes |{\rm n} \rangle$,
 since there are complicated exchanges of quarks and gluons between the two composite
 particles. 
 
 The spatial extent of the NN interaction in QCD is short ranged and is 
 exponentially suppressed beyond the distance $R > 2 $ fm.
 Therefore, the spatial part of the BS wave function in the 
  ``outer region" ($r > R$)
  satisfies the Helmholtz equation below the pion production threshold,
   $(\nabla^2 + k^2 )\psi_{\alpha \beta}(\vec{r}) =0$,
 up to an exponentially small correction.
 Then we can define the non-local potential $U$ from $\psi$ and $k^2$ measured
  on the lattice:
 \beq
 \label{eq:QCD_Schroedinger}
 \! \! \! \! \! \! (E- H_0) \psi_{\alpha \beta,E}(\vec{r}) 
 &=&  \int U_{\alpha \beta; \gamma \delta}(\vec{r},\vec{r}') \psi_{\gamma \delta,E}(\vec{r}') d^3r' , \\
 \label{eq:QCD_nonlocal-potential}
\! \! \! \! \! \!  U_{\alpha \beta; \gamma \delta}(\vec{r},\vec{r}') 
 &=&     V_{\alpha \beta; \gamma \delta}(\vec{r}, \vec{v}) \delta(\vec{r}-\vec{r}'), 
\eeq  
 where $\vec{v} (= \vec{p}/\mu = -i \nabla/\mu)$ is the velocity operator.
 To make a formal resemblance with the 
 non-relativistic case, we have introduced  the ``effective center of mass  energy",
 $E=k^2/(2\mu)=k^2/m_N$ and the ``free Hamiltonian", $H_0= -\nabla^2/m_N$. 
  By construction, the solution of Eq.(\ref{eq:QCD_Schroedinger})  with 
  $U_{\alpha \beta; \gamma \delta}(\vec{r},\vec{r}')$  
   extrapolated to $L \rightarrow \infty$
  reproduces the correct BS wave function in the asymptotic region, and hence the
   phase shifts and binding energies of the two-nucleon system.

The simplest interpolating operators  for the nucleon $N=(p,n)$ 
 in terms of the quark field $q(x)$ would be 
$  N_\beta(x) =  \varepsilon_{abc} \left( q_a(x) C \gamma_5 q_b(x) \right)
  q_{c\beta}(x)$, with 
  $a$, $b$ and $c$ being color indices and $C$ being the 
 charge conjugation matrix. 
 Such a local operator is
   most convenient for relating  the BS wave function to the 
   four-point Green's function and the scattering 
   observables at $L \rightarrow \infty$. 
  Closely related formulation was given  
  long time ago by Nishijima, Zimmermann and Hagg
  who derived the generalized  reduction formula   for 
  local composite fields \cite{NZ}. 
 
  In principle,  one may choose any composite operators
  with the same quantum numbers as the nucleon to define the BS wave
  function.   Different  operators
  give different BS wave functions and different NN potentials, although
  they lead to the same observables.
  This is quite analogous to the  situation in quantum mechanics
  where the unitary transformation 
  of the wave function changes the structure of the 
   potential while  the observables are not modified. 
   A theoretical advantage of our approach based on lattice QCD is that
  we can unambiguously trace the one-to-one correspondence between
  the NN potential and the interpolating operator in QCD as we mentioned.

 The general form of the non-local NN potential $U$ or equivalently
  the velocity dependent NN potential $V$ in  Eq.(\ref{eq:QCD_nonlocal-potential})
   in the two-component 
 spinor space has been classified by Okubo and Marshak \cite{okubo}.
The leading order (LO) and the next-leading-oder (NLO) terms of the 
 the velocity expansion of $V(\vec{r},\vec{v})$ reads \cite{TW67} 
\beq
\label{eq:OM-pot-2}
V & =&
   V_C(r) + V_T(r) S_{12} +  V_{LS}(r) {\bL} \cdot {\bS}  +{O}(\vec{v}^2), \\
 & =&
   V_0(r)
  +V_\sigma(r)(\bsigma_1 \cdot \bsigma_2)
  +V_\tau(r)(\btau_1 \cdot \btau_2) 
   +V_{\sigma\tau}(r)
   (\bsigma_1 \cdot \bsigma_2)
   (\btau_1 \cdot \btau_2)
   \nonumber \\
 &&
  +\left[ V_{T0}(r)
  +V_{{T}\tau}(r)(\btau_1 \cdot \btau_2)\right] S_{12} \nonumber \\
 &&+\left[ V_{LS0}(r)
  +V_{{LS}\tau}(r)(\btau_1 \cdot \btau_2) \right] {\bL} \cdot {\bS}
  +{O}(\vec{v}^2), 
 \label{eq:orderP}   
\eeq
where $V_C$ and $V_T$ are  LO ($O(\vec{v}^0)$) 
terms, while $V_{LS}$ is a NLO ($O(\vec{v})$) term.
 On the lattice, it is relatively  unambiguous
  to extract information for the orbital angular momentum states
  $\ell=0,1,2,3 = {\rm S, P, D, F}$
using  the irreducible representations of the cubic group \cite{luescher}. 
Then, at most 16 independent (14 diagonal and 2 off-diagonal) matrix elements 
of the potential
 are obtained, so that 8 unknown  LO and NLO terms 
  in Eq.(\ref{eq:orderP})  can be extracted in two different ways.

\section{Central and tensor forces from lattice QCD}
\label{sec:central}

To define the BS wave function on the lattice, 
 we start from the four-point correlator,
\begin{eqnarray}
\label{eq:4-point}
 {\cal G}_{\alpha \beta} 
= \left\langle {\rm vac}
   \left|
    n_\beta(\vec{y},t)
    p_\alpha(\vec{x},t)
    \overline{\cal J}_{pn}(t_0;J^P)
   \right| {\rm vac} 
  \right\rangle    \rightarrow 
 A_0 \ \psi_{\alpha \beta}(\vec{r};J^P)\  {\rm e}^{-E_0(t-t_0)}
 \ \ \ (t \gg t_0),
\label{eq:BSamp}
\end{eqnarray}
where $A_0$ is an  $r$-independent constant. 
The states created by the source  $\overline{\cal J}_{pn}$
 have the conserved quantum numbers, 
   $(J,J_z)$ (total angular momentum and its z-component) and $P$ (parity).
 For studying the nuclear force in the  $J^P=0^+$ ($^1{\rm S}_0$) channel and
  the  $J^P=1^+$ ($^3{\rm S}_1$ and $^3{\rm D}_1$) channel, we adopt
   a wall source with 
    the Coulomb gauge fixing at $t=t_0$.
 The BS wave function in the orbital S-state is  defined  with 
 the projection operator for the orbital angular momentum ($P^{(\ell)}$) 
 and that for the spin ($P^{(s)}$) as  
 $\psi(r; ^1{\rm S}_0) = P^{(\ell =0)}  P^{(s=0)}   \psi(\vec{r};0^+)$ and 
 $\psi(r; ^3{\rm S}_1) = P^{(\ell =0)}  P^{(s=1)}   \psi(\vec{r};1^+)$.

The asymptotic momentum $k$ for the S-states is  obtained 
 by fitting the BS wave function $\psi(\vec{r})$ with the Green's 
 function $G(\vec{r}; k^2)$ in a
finite and periodic box 
 satisfying $(\nabla^2 + k^2) G(\vec{r}; k^2) = - \delta_{\rm lat}(\vec{r})$ with
 $\delta_{\rm lat}(\vec{r})$ being the periodic delta-function. 
  The fits are  performed outside the range  of the NN interaction
   determined by  $\nabla^2\psi(\vec{r})/\psi(\vec{r})$ \cite{ishizuka}.
 The NN scattering lengths  for the S-states can be deduced from
the standard L\"{u}scher's formula \cite{luescher}.

  In the LO of the velocity expansion, 
only the central potential $V_C(r)$  and  the tensor potential  
$V_T(r)$ are relevant: The central potential  acts
separately on the S and D  components, while  the  tensor potential
 provides a coupling between these two. Therefore,
  we consider a coupled-channel
 Schr\"odinger equation in the  $J^P=1^+$ channel \cite{Ishii:2009zr}:
\beq
  \bigl( H_0 + V_{C}(r) + V_{T}(r) S_{12} \bigr)
  \psi(\vec{r};1^+)
  =
  E
  \psi(\vec{r};1^+).
  \label{schrodinger.eq.one.plus}
\eeq 
Projections to the S-wave and D-wave components are obtained as
  $ \psi_{\alpha \beta}(r; ^3{\rm S}_1)  \equiv 
  P^{(\ell =0)}  \psi_{\alpha \beta}(\vec{r};1^+)$ and   
 $\psi_{\alpha \beta}(r; ^3{\rm D}_1)  \equiv 
 (1- P^{(\ell =0)} ) \psi_{\alpha \beta}(\vec{r};1^+)$. 
  In the LO of the velocity expansion, it is sometimes useful to define  
 the  ``effective" central potential $V_{C}^{\rm eff}(r)$ \cite{Ishii:2006ec}: 
$  V_{C}^{\rm eff}(r) = 
  E + {1\over m_{N}}{{\nabla}^2\psi(r)\over \psi(r)} $. 
 Note  that $V_{\rm  C}^{\rm eff}(r)$  in the $ ^3{\rm S}_1$ channel
 contains the  effect  of 
$V_T(r)$ implicitely as higher order effects through the process such as
 $^3{\rm S}_1 \rightarrow ^3{\rm D}_1 \rightarrow ^3{\rm S}_1$.

\section{Numerical results in quenched QCD}
\label{sec:numerical} 

   In the quenched simulations, 
   we  employ  the  standard  plaquette  gauge action on  a $32^4$  lattice  with
 the bare QCD coupling constant $\beta = 6/g^2 =  5.7$.
 The corresponding lattice spacing  is 
 $1/a=1.44(2)$ GeV  ($a\simeq 0.137$ fm)  
 determined  from the $\rho$ meson mass in the chiral limit.
  The physical size of our lattice then reads $L\simeq 4.4$ fm.
  We adopt the standard Wilson quark action  with the 
  hopping parameter $\kappa=0.1640, 0.1665, 0.1678$,
  which correspond to $ m_{\pi} \simeq 731, 529, 380$ MeV, respectively.
 The  periodic  boundary  condition is  imposed on  the quark
  fields along the spatial direction, while
  the Dirichlet boundary  condition is imposed in the
temporal direction  at the time-slice  $t=0$.  The wall source is placed on
the time-slice at $t_0/a \equiv 5$ with the Coulomb gauge fixing at $t=t_0$.
 The lowest effective c.m. energy $E$  in the above setup ranges from 
  $-0.4$ MeV to $-1$ MeV. Note that $E$ for  
  scattering states can be negative in a finite box.

\subsection{Central and tensor forces in the $^3 {\rm S}_1$ channel} 

\begin{figure}[t]
\begin{center}
\includegraphics[width=6cm,angle=-90]{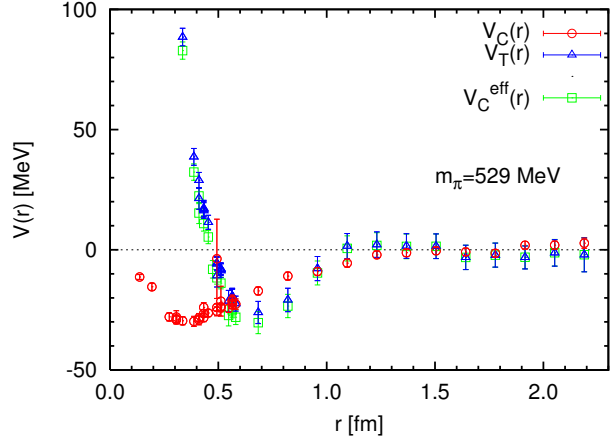}
\end{center}
\caption{The  central potential $V_C(r)$  and the tensor
potential 
$V_T(r)$ obtained from the $J^P=1^+$ BS wave function at
$m_{\pi}=529$ MeV.   }
\label{fig:tensor.force}
\end{figure}

 Shown in \Fig{fig:tensor.force}  is the central
potential  $V_C(r)$ and  tensor potential  $V_T(r)$  together with
effective central potential $V_C^{\rm eff}(r)$ in the $^3 {\rm S}_1$ channel
 obtained in the LO velocity expansion.
 In the real world,
 $V_{\rm  C}^{\rm eff}(r)$ is expected to acquire sufficient
attraction from the tensor force. This is the reason why
 bound deuteron exists in the $^3{\rm S}_1$ channel while the bound
  dineutron does not exist in the $^1{\rm S}_0$ channel. 
 Now, we see from \Fig{fig:tensor.force}  that the difference between 
$V_C(r)$ and 
$V_C^{\rm eff}(r)$  is  still small in our quenched simulations due to 
 relatively large quark masses. 

\begin{figure}[t]
\begin{center}
\includegraphics[width=6cm,angle=-90]{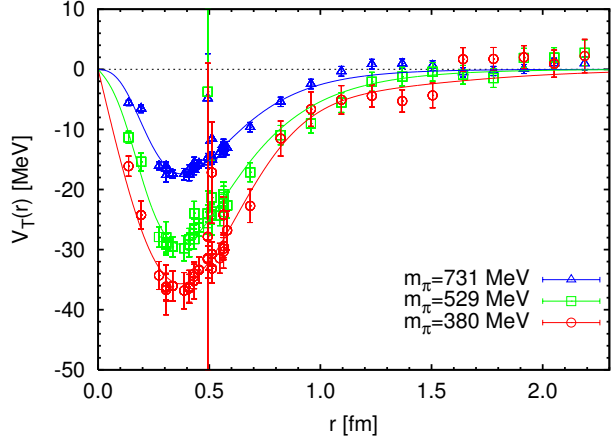}
\end{center}
\caption{Quark mass dependence of tensor potential.
The lines are
 the four-parameter fit
  using the one-$\rho$-exchange +  one-pion-exchange with Gaussian form factors. }
\label{fig:tensor.force.2}
\end{figure}

 The  tensor potential $V_T(r)$  in \Fig{fig:tensor.force}
 shows that it is negative for the whole range of $r$ with a minimum
 at short distance below  $0.5$ fm.  If the long range part of the tensor force
  is dominated by the one-pion exchange as expected from the meson theory,
  $V_T(r)$ could be rather sensitive to the change of the
  quark mass.  As shown in \Fig{fig:tensor.force.2},
   it is indeed the case:    Attraction of $V_T(r)$ is
    substantially enhanced as the quark mass decreases. 
  For practical 
 applications in nuclear physics,
 it is more useful to parametrize the lattice results 
 by known functions. We have tried such a fit for 
 $V_T(r)$ under the assumption of 
  the one-$\rho$-exchange + one-pion-exchange with Gaussian form factors:
 $V_T(r) = b_1 (1- e^{-b_2 r^2})^2
\left( 1 + \frac{3}{m_{\rho}r} + \frac{3}{(m_{\rho}r)^2} \right)  \frac{e^{-m_{\rho}r} }{r} 
 +  b_3 (1- e^{-b_4 r^2})^2 \left( 1 + \frac{3}{m_{\pi}r} + \frac{3}{(m_{\pi}r)^2} \right)
  \frac{e^{-m_{\pi}r} }{r}$. The results are
  shown by the solid lines in \Fig{fig:tensor.force.2}.

\subsection{Velocity dependence of the potential}
\label{sec:vel-dep}

\begin{figure}[t]
\begin{center}
\includegraphics[width=8.5cm]{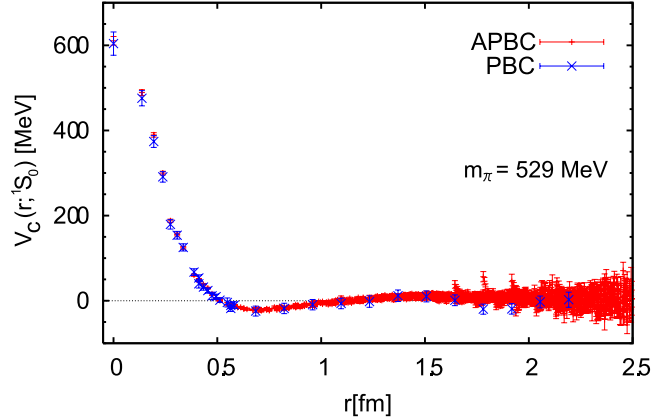}

\end{center}
\caption{The central NN potentials for the $^1{\rm S}_0$
 state with APBC (red bars) and PBC (blue crosses) in quenched QCD at $t/a=9$.
}
\label{v-dep}
\end{figure}

  So far we have considered the potential determined from the 
  lattice data taken almost at zero effective c.m. 
    energy $E \simeq 0$ MeV.
  If the local potential  determined from the other energies have 
  different spatial structure, it is an indication of the 
  velocity dependent terms.
  Such a velocity dependence  has been recently studied by changing the spatial 
   boundary condition of the quark field from the periodic one
  to the anti-periodic one  \cite{murano}:
  On a  $32^3 \times 48$ lattice with the lattice spacing $a=0.137$ fm, 
  2000 gauge configurations are accumulated.
  The minimum momentum is given by $\vec{p}_{\rm min} = (\pi,\pi,\pi)/(32a)$, 
  which leads to $|\vec{p}_{\rm min}| \simeq 240$ MeV and
 $E  \simeq 50$ MeV.  In Fig.\ref{v-dep}, 
 the central NN potential for the $^1{\rm S}_0$ 
state with  APBC ( $E\simeq 50$ MeV) is plotted as a function of $r$ at $t/a=9$, 
together with the one with PBC ($E \simeq 0$). 
Fluctuations of the data with APBC at large distances ($ r > 1.5$ fm) are mainly 
caused by contaminations from excited states, together with statistical noises.  
 The potential at $ r < 1.5$ fm, on the other hand, is less 
affected by such contamination.
As seen from Fig. \ref{v-dep}, the NN potentials are almost 
identical between $E\simeq 0$ and $E\simeq 50$ MeV.  Namely, the 
   non-locality of the potential with our choice of the interpolating operator
  is small and the LO potentials shown in the present paper can be used
   in the energy region at least up to $E \sim 50$ MeV without 
   significant  modifications.

\section{Numerical result in (2+1)-flavor QCD}
\label{sec:full-qcd}

\begin{figure}[t]
\begin{center}
\includegraphics[width=6cm,angle=-90]{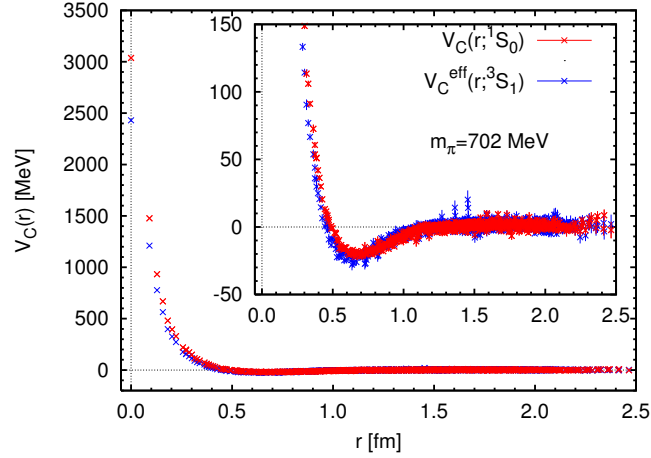}
\end{center}
\caption{Central and effective central potentials in
 (2+1)-flavor lattice  QCD simulations with $m_{\pi}=702$ MeV.
}
\label{fig:full}
\end{figure}

To compare our results with empirical  data, a key role is played by a
full QCD calculation on a large volume with a smaller quark mass.
The PACS-CS collaboration is generating (2+1)-flavor gauge configurations  
 by employing the Iwasaki gauge
action at  $\beta=1.90$ on  $32^3\times 64$ lattice  and the $O(a)$-improved
Wilson  quark  (clover)  action  with  a  non-perturbatively  improved
coefficient $c_{\rm SW}=1.715$ \cite{Kuramashi:2008tb}.
The  lattice   scale  is   determined  by   $m_{\pi}$,   $m_{K}$  and
$m_{\Omega}$, which leads to $a\simeq 0.091$ fm.
 Thus,  the spatial extension amounts  to $L \simeq 2.90$ fm. 
The  periodic  boundary
condition is imposed  along the  spatial direction, while
 the Dirichlet boundary condition on the
time-slice $t=32$ is imposed along the temporal direction.
 The wall source on the time-slice is located at 
$t=0$ with the Coulomb gauge fixing.
 \Fig{fig:full} shows  the full  QCD  results of  the
central   force for $m_{\pi} \simeq 702$ MeV:
$V_{\rm C}(r; ^1S_0)$ and $V_{\rm C}^{\rm eff}(r; ^3S_1)$ are obtained
from  BS   wave  functions  on   the  time-slices  $t=8$   and  $t=9$,
 respectively, where  the ground state saturations  are achieved within
 error bars. Similar  to the quenched  results,
 a repulsive core surrounded by an attractive well can be seen in full QCD.

\section{Concluding remarks}
\label{sec:summary}

 We have discussed  
  the basic notion of the nucleon-nucleon potential and its field-theoretical 
  derivation from the  equal-time Bethe-Salpeter amplitude in QCD. 
  By construction, the non-local potential defined through the projection of the
   wave function to the interaction region (the inner region)   correctly reproduces the 
   asymptotic form of the wave function in the region beyond the 
   range of the nuclear force (the outer region). Thus the observables such as the 
   phase shifts and the binding energies can be
   calculated after extrapolating the potential to the infinite volume limit.
   Non-locality of the potential can be taken into account successively by
    making its velocity expansion, which introduces the velocity-dependent
     local potentials. The leading order terms  of such velocity expansion 
     are the central potential and the tensor potential, and the next-to-leading order
   term is the spin-orbit potential.  
           
  As an exploratory study,
   we carried out  quenched lattice  QCD simulations of the two-nucleon system
  in a spatial box of the size (4.4 fm)$^3$ with the pion mass
   $m_{\pi} = 380, 529, 731$ MeV. 
   The NN potential calculated on the lattice at low energy 
  is found to have  all the characteristic features 
  expected from the empirical NN potentials obtained from the experimental
   NN phase shifts, namely  the repulsive core surrounded by the 
  attractive well for the central potential. As for the tensor potential obtained by the  
  coupled channel treatment of the $^3{\rm S}_1$-state
  and the $^3{\rm D}_1$-state,
 appreciable attraction at long and medium distances is found.
   Phenomenological fit of the tensor potential
       strongly suggests the existence of the one-pion-exchange
      in its long range part.

  There are a number of  directions to be investigated
   on the basis of our approach:
\begin{itemize}
  \item[1.] Determination of the velocity dependence  is
   important in deriving the NN potentials useful for the wide
    range of scattering energies.
    As mentioned in Sec.\ref{sec:vel-dep},
  studies along this line using the anti-periodic boundary condition
   in the spatial direction have been already started  \cite{murano}.  
  \item[2.] To derive the realistic  NN potentials on the lattice,
    it is necessary to   carry out full QCD simulations with  dynamical quarks.
   As mentioned in Sec.\ref{sec:full-qcd}, studies along this line  with the use of the  
    (2+1)-flavor QCD configurations generated by PACS-CS
    Collaboration  are currently under way \cite{Ishii:2009zr}.
  \item[3.] The hyperon-nucleon (YN) and hyperon-hyperon (YY)
   potentials are essential for understanding the properties
    of hyper nuclei and the hyperonic matter inside the
     neutron stars.   
   Recently,  studies of  the YN potential in quenched QCD \cite{Nemura:2008sp}
   and in full QCD \cite{Nemura:2009kc} have  been started.
  \item[4.] The three-nucleon force is thought to play important roles in
 nuclear structures and  in the equation of state 
 of  high density matter \cite{Pieper:2001ap}. 
  Since the experimental information is
 scarce, simulations of the three nucleons  on the lattice  may
 lead to the first principle determination of the three-nucleon 
 potential in the near future.   
\end{itemize}


\end{document}